\documentclass[aps,pre,twocolumn,groupedaddress,showpacs]{revtex4}

\usepackage{epsfig,amsmath,amsbsy,amssymb}

\usepackage{bm}

\usepackage{color,colordvi}
\definecolor{Red}{rgb}{0.9,0.0,0.1}

\begin{document}

\title{Cooperative dynamics  of microtubule ensembles:
  Polymerization forces and rescue-induced oscillations}

\author{Bj\"orn Zelinski and Jan Kierfeld}
\affiliation{Physics Department, TU Dortmund University, 
44221 Dortmund, Germany}

\date{\today}
\begin{abstract}
We investigate the cooperative dynamics of  
an ensemble of $N$ microtubules
growing against an  elastic barrier.
Microtubules undergo so-called catastrophes,
which are abrupt stochastic transitions from a growing
to a shrinking state, and rescues, which are transitions 
back to the growing state.
Microtubules can exert pushing or polymerization forces on an 
obstacle, such as an elastic barrier if the 
growing end is in contact with the obstacle.
We use  dynamical mean-field theory and  
stochastic simulations to analyze a  model where 
each microtubule undergoes catastrophes and rescues
and where microtubules interact by force sharing.
For zero rescue rate, cooperative growth 
terminates in a collective catastrophe. The 
 maximal polymerization force before catastrophes 
 grows linearly with $N$  for 
small $N$ or  a stiff elastic barrier, in agreement with 
available experimental results,
whereas it crosses over to a logarithmic dependence for larger $N$
or a soft elastic barrier.
For a nonzero rescue rate and a soft elastic barrier, the dynamics becomes 
 oscillatory  with  both
collective catastrophe and rescue events, which are part of 
 a robust limit cycle.
Both the  average and maximal polymerization forces then
grow  linearly with  $N$, and we investigate
their dependence on tubulin on-rates and rescue rates,
which can be involved in cellular regulation mechanisms.
We further investigate the robustness of the 
collective catastrophe and rescue 
oscillations with respect to different catastrophe models. 
\end{abstract}


                            
\maketitle

\section{Introduction}

Microtubules (MTs) are long and stiff filamentous proteins, which assemble 
and disassemble from tubulin dimers and serve various functions 
in the cytoskeleton: MT stiffness plays an important role 
in cytoskeletal mechanics but they also serve as ``tracks'' for 
intracellular transport by molecular motors. 
Finally, MTs exhibit an unusual polymerization 
dynamics, which is essential  for 
spatial organization and remodelling processes in the cytoskeleton
\cite{Desai1997}.

Polymerizing MTs can also generate pushing  forces 
within the cell,
 because  MTs grow by tubulin insertion also
 in the presence  of an obstacle which exerts 
an opposing force on tubulin dimers inserted at the 
 growing MT tip. 
The  opposing force  is transmitted onto the MT, as has been 
demonstrated in single-MT buckling experiments in front of a 
solid wall \cite{DY97}.
The force slows down further MT growth because 
monomer insertion against a force involves additional mechanical 
work. Growth finally terminates under a maximal polymerization 
force, which is typically in the pN range \cite{DY97}.

An important and unique 
 feature of the MT polymerization dynamics is 
their so-called {\em dynamic instability}.
The dynamic instability gives rise to
phases of fast shrinking, which  stochastically 
interrupt polymerization phases \cite{Mitch1984}. 
Each  phase of of fast shrinking is initiated by a 
catastrophe event and terminated by a rescue event.
This complex dynamic behavior is central to rapid remodelling 
in the cytoskeleton but also affects the  polymerization force.

Catastrophe and rescue events of MTs are associated
with guanosine triphosphate (GTP) hydrolysis within the MT \cite{Mitch1984}.
Each tubulin dimer consists of an $\alpha$- and $\beta$-tubulin and 
contains two GTP binding sites. 
Tubulin dimers assemble into polar  MTs with a fast-growing plus end. 
GTP-tubulin dimers attach to the plus end  
with both binding sites containing GTP.
The GTP in the $\alpha$-tubulin is hydrolyzed to 
 guanosine diphosphate (GDP) within the MT. 
This process leads to the formation of a GTP cap at the growing 
plus end,
whereas the remaining MT consists of GDP-tubulin. The size of the GTP cap
fluctuates in time and depends on the interplay of 
GTP-tubulin on-rate and hydrolysis rates within the MT. 
The GTP cap stabilizes the MT structure mechanically, and 
catastrophes are triggered by the loss of the GTP cap. 
The catastrophe rate therefore is  related to the 
hydrolysis dynamics within the MT and given by the  
first passage rate  to a state with vanishing GTP cap.
There are different models describing this process 
\cite{Flyv1994PRL,Flyv1996PRE,Janson2003}, 
which will be discussed in Sec. \ref{sec_catmodels}
in more detail. The dynamic instability also limits the 
 ability of MTs to generate polymerization 
forces \cite{Zelinski2012}.

In a living cell, MTs often cooperate in order 
to generate higher forces. 
Cooperative MT polymerization forces  play an 
important  role during mitosis 
in generating forces necessary for chromosome 
separation \cite{McIntosh2002}.
A strong cooperativity in the dynamics  of MTs
is also relevant in processes 
 regulating the cell length, as has 
been reported  in Ref.\  \cite{Picone2010} for animal cells.
 MTs also cooperate 
in  the formation of cell protrusions, for example,
in neuronal growth \cite{Poulain2010}. 
Therefore, it is important to develop models and a theoretical 
framework for the dynamics of MT ensembles under force, which can 
describe cooperative  effects in the polymerization dynamics and 
quantify the cooperatively  generated  MT forces.

Only recently, has it become possible to study 
force generation by MT ensembles {\em in vitro} using MT bundles 
 growing against an elastic force, 
which was realized by an optical trap \cite{Laan2008}.
Experimental conditions were such that no rescue events occurred.
The experiments showed {\it collective catastrophes} of the whole MT bundle:
Before a collective catastrophe, 
 a large fraction of the MT ensemble is  growing 
cooperatively against the elastic force;
growth often terminates in a  catastrophe of the whole bundle where 
 all pushing  MTs 
nearly simultaneously undergo a catastrophe.
In the experiments in Ref.\ \cite{Laan2008}, 
the  maximal polymerization force that is reached 
before a collective catastrophe 
was measured to  grow linearly with the number 
$N$ of MTs in the ensemble.

In Ref.\ \cite{Laan2008}, the experimentally observed 
collective catastrophes, and the linear $N$ dependence, could be 
reproduced in simulations, where 
polymerizing MTs grow against an elastic force and 
 interact by force sharing. 
A theory explaining the characteristic linear
dependence of the maximal polymerization force 
on the number $N$ of MTs  is, however, lacking. 
The simulations in Ref.\ \cite{Laan2008}
 included  renucleation of MTs after complete
depolymerization, which  led to oscillations in the growth dynamics
of the MT ensemble.
The simulations did not include rescue events, which can be relevant 
{\em in vivo}.

In the present article, we will use a 
model for the collective dynamics of $N$ MTs growing  against an 
elastic  barrier which is very similar to the model that has been used 
in the simulations in Ref.\ \cite{Laan2008}.
As in Ref.\ \cite{Laan2008}, 
the ensemble of MTs grow against an elastic force, which 
 models the optical traps used in the experiments  or 
the elastic cell cortex {\em in vivo}.
As in Ref.\ \cite{Laan2008}, 
the most important features of the model are that
MTs only interact by 
 force sharing between  the cooperatively pushing  leading 
MTs and that the  single-MT catastrophe rate increases exponentially 
with its load force.
In contrast to  Ref.\ \cite{Laan2008}, 
we will not consider renucleation of shrinking MTs but 
include  rescue events into the model, which 
are an essential part of MT dynamics, and study their influence 
on the cooperative MT dynamics.

Furthermore, we also 
 develop a dynamical mean-field theory, which provides a theoretical 
framework to describe  the cooperative MT dynamics both in the 
absence and presence of rescue events. It allows us 
to extract the relevant control parameters, such as 
tubulin on-rate, rescue rate, and MT number, and to 
investigate  their influence on 
dynamics and force generation. 
It is  shown that, apart from collective catastrophes, also 
collective rescue events emerge if rescue events are 
taken into account in  the single-MT dynamics. 
The resulting interplay between collective catastrophes
and collective rescue events gives rise to an
oscillatory growth dynamics of the entire MT ensemble. 
The  dynamical mean-field theory 
successfully describes this oscillatory dynamics as a robust limit 
cycle.

The dynamical mean field theory
 allows us to calculate  the cooperative polymerization
force generated by the MT ensemble and its dependence 
 on the  MT number $N$,  the stiffness of the barrier, 
and the MT growth parameters.
 Furthermore, we  use the 
dynamical mean field theory to 
investigate the robustness of our results 
against variations of the catastrophe model.
This is an important question as at least two 
different catastrophe models
have been put forward in the literature and have been shown 
to describe experimentally available data on single-MT catastrophe 
rates. Robustness of our results show that 
details of the catastrophe models are not essential for 
force generation by 
ensembles of MTs but only rather general features, such as the
exponential increase of the catastrophe rate with force,
are relevant. 
We corroborate all our mean-field results by 
microscopic stochastic simulations of the full MT 
ensemble dynamics.

The paper is structured as follows. In Sec. \ref{sec_model}, we  
introduce the model describing the stochastic growth dynamics 
of an ensemble of $N$ MTs growing against an elastic barrier. 
We, first, outline the stochastic growth model for a single MT, 
 the catastrophe models used throughout the work, and,
finally, the coupling between the $N$ MTs in the ensemble 
dynamics via force sharing between leading MTs.
In Sec. \ref{sec_parameters} we describe our choice of 
model parameters and describe  the simulation.

We then develop the  dynamical mean-field  theory for this 
stochastic model. 
In Sec. \ref{no_rescue}, we start with the case of 
zero rescue rate, which has direct applications to the experiments 
by Laan {\it et al.} \cite{Laan2008} and for which 
 the  cooperative MT dynamics is conceptually 
simpler to understand because there are only collective 
catastrophes.
For zero rescue rate, we discuss the maximal polymerization force and 
find a linear $N$ dependence for small $N$ or stiff barriers with a 
crossover to a logarithmic dependence for large $N$ or soft barriers.
The results agree with the experimental findings 
of  Ref.\ \cite{Laan2008}.

In Sec. \ref{sec_rescue},
 we introduce the dynamical  mean-field theory for 
 the full problem in the presence of rescue events,
where the cooperative MT dynamics exhibits both collective catastrophes
and collective rescue events. 
For nonzero rescue rate and a soft barrier,
we find stable collective catastrophe and rescue oscillations with  
maximal and average polymerization forces growing linearly with $N$.
We show that our theory is applicable to different catastrophe 
models for single MTs and that our main findings are robust 
for catastrophe models with catastrophe rates exponentially increasing 
with force.

Throughout the article, we show that all theoretical results are 
in agreement with 
 microscopic stochastic simulations.

\section{Model for cooperative dynamics of MT ensembles}
\label{sec_model}

\subsection{Single MT model}

The dynamic instability
causes the MT plus end to switch stochastically  between 
a growing (+) and a shrinking (-) state \cite{Dogt1993}. 
In the  growing state, 
GTP-tubulin dimers (called {\em monomers} in the following) 
 attach and detach
with  rates $\omega_{\text{on}}$ and  $\omega_{\text{off}}$, respectively, 
to one of the 13 protofilaments. 
The MT  growth velocity in the growing state 
and in the absence of external forces 
is   $v_+=d(\omega_{\text{on}}-\omega_{\text{off}})$
[$v_+=1\ldots 5 \times 10^{-8}\;{\rm m/s}$],
where  $d\simeq 8\,{\rm nm}/13$ is the  effective monomer size.

Under force, the MT growth velocity becomes force 
dependent 
\begin{equation}
   v_+(F) =d \left[ \omega_{\text{on}} e^{-F/F_0}-\omega_{\text{off}} \right]
\label{v+F}
\end{equation}
with a characteristic force $F_0 \equiv k_BT/d \simeq 7\;{\rm pN}$.
For simplicity, 
we assume that force only affects the on-rate of tubulin monomers. 
Experimental measurements of the force-velocity relation in 
Ref.\ \cite{DY97} gave a significantly 
smaller value $F_0 \simeq 2\;{\rm pN}$.

\subsection{Dynamic instability and  catastrophe models}
\label{sec_catmodels}

The dynamic instability of MTs is  associated 
with the loss of the stabilizing GTP cap  because 
of hydrolysis within the MT \cite{Mitch1984}.
At the catastrophe rate $\omega_c$ the MT loses its GTP-cap in the 
growing state and  switches into a state of fast shrinking with 
a large shrinking velocity $v_-$ ($\simeq 3\times 10^{-7}\;{\rm m/s}$). 
With the rescue rate $\omega_r$ the MT switches back from 
the shrinking  into the growing state.

The catastrophe rate is  related to the 
hydrolysis dynamics within the MT and given by the  
first passage rate  to a state with vanishing GTP-cap.
The resulting catastrophe rate has been discussed based on a 
model for {\em cooperative} hydrolysis of GTP-tubulin
by Flyvbjerg {\it et al.}
\cite{Flyv1994PRL,Flyv1996PRE}.
Similar cooperative 
models have been proposed 
for the hydrolysis dynamics in filamentous 
actin \cite{Kier2009,Kier2010}. 
In a cooperative model, hydrolysis proceeds  by 
 a combination of both {\em random}  and {\em vectorial} 
mechanisms. In a random mechanism, 
the hydrolysis rate  $r$ per length 
($\simeq 3.7\times 10^6\, {\rm m}^{-1}{\rm s}^{-1}$)
of GTP-tubulin is independent of the position of 
the GTP-monomer in the MT.
In a vectorial mechanism, only GTP-monomers are hydrolyzed 
which have already hydrolyzed 
GDP-monomers at one neighboring site.
 This results in a directed motion of 
 GTP- and GDP-tubulin interfaces with mean velocity  
$v_h$   ($\simeq 4.2\times 10^{-9}\;{\rm m/s}$).
The inverse catastrophe rate $\omega_c^{-1}$ can be obtained as
mean first passage time to a state with a vanishing GTP cap.
For a cooperative model,  the  exact analytical result for $\omega_c$ 
has been obtained in Refs.\ \cite{Flyv1994PRL,Flyv1996PRE}
as implicit function of the growth velocity  
$v_{+}$ and the hydrolysis parameters 
 $v_h$ and $r$.
 The exact dimensionless catastrophe rate 
$\alpha \equiv \omega_{c} D^{-1/3}r^{-2/3}$ is 
given by the smallest solution of 
\begin{equation}
   Ai'(\gamma^2-\alpha) = -\gamma Ai(\gamma^2-\alpha)
\label{Ai}
\end{equation}
with $\gamma \equiv vD^{-2/3}r^{-1/3}/2$,
where $v\equiv v_+-v_h$ and $D\equiv (v_{+}+v_h)d/2$ 
[$Ai'(x) \equiv dAi(x)/dx$].
Here $Ai$ denotes the first Airy function \cite{Abramowitz}. 
We use 
a numerical implementation of this
 exact analytical result for $\omega_c$
in simulations and mean-field calculations:
We solve (\ref{Ai}) to calculate the function 
$\alpha = \omega_{c} D^{-1/3}r^{-2/3}$ as a function of $\gamma$ numerically.
From this numerical solution  we obtain the function 
$\omega_c = \omega_c(v_+)$,  for which we generate 
an accurate interpolating polynomial of high order.
This polynomial is used in the simulation to calculate 
catastrophe rates as a function  of (force-dependent) 
growth velocity, $v_+(F)$.
The remaining hydrolysis parameters $v_h$ and $r$ are
fixed during the simulation.

\begin{figure}
 \begin{center}
 \epsfig{file=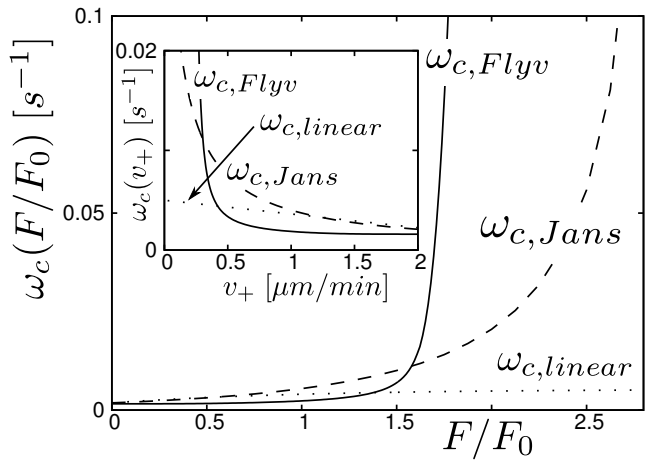,width=0.47\textwidth}
 \caption{\label{fig:omegac} The catastrophe rate $\omega_c$ (in ${\rm
     s}^{-1}$) as a function of the load force $F$ (in units of $F_0$)
   for the Flyvbjerg model (with $F_0 \sim 7\;{\rm pN}$ and for
   $\omega_{\text{on}} = 70\; {\rm s}^{-1}$), the linear catastrophe
   model (with $\tilde{a} = 0.005\;{\rm s^{-1}}$,$\tilde{b} = 8 \times
   10^5\;{\rm m^{-1}}$, $F_0 \sim 7\;{\rm pN}$, and
   $\omega_{\text{on}} = 70\; {\rm s}^{-1}$) and the Janson model
   (with $F_0 \sim 0.8\;{\rm pN}$).  Inset: The catastrophe rate
   $\omega_c$ (in ${\rm s}^{-1}$) as a function of the growth velocity
   $v_{+}$ (in $\mu m/\text{min}$) for the three different catastrophe
   models.
}
 \end{center}
\end{figure}

The catastrophe rate $\omega_c$ becomes force dependent 
via the force-dependence (\ref{v+F}) of the growth velocity.
The resulting catastrophe rate is a nonlinear and 
monotonically increasing function
of the force $F$ (see Fig.\ \ref{fig:omegac}), which increases 
exponentially  above the characteristic force $F_0$.
We use the theoretical  value $F_0 = k_BT/d \simeq 7\;{\rm pN}$ 
for the  model by Flyvbjerg {\it et al.} in the following.

We  will investigate whether collective catastrophe and rescue 
oscillations are robust with respect to the  catastrophe model.
Different catastrophe models
have been proposed  and have been shown 
to describe experimentally available data on single-MT catastrophe 
rates.
One  alternative phenomenological 
catastrophe model has been proposed by 
Janson {\it et al.} based on experimental data for the 
 inverse catastrophe rate, i.e., 
the average  catastrophe time $\tau_c = 1/\omega_c$ \cite{Janson2003}.
The experimental data show  that $\tau_c$ 
 increases linearly with the growth velocity $v_+$ such 
that the catastrophe rate is  given by 
\begin{equation}
   \omega_c = \frac{1}{a+b v_+}.
\label{Janson}
\end{equation}
with   $a\simeq 20\;{\rm s}$ and 
$b\simeq 1.4\times 10^{10}\;{\rm s}^2m^{-1}$ \cite{Janson2003}.

Also within the catastrophe model by Janson {\it et al.}, the 
catastrophe rate $\omega_c$ becomes force dependent 
via the force dependence (\ref{v+F}) of the growth velocity, and 
 the resulting catastrophe rate is a nonlinear and increasing function
of the force $F$ (see Fig.\ \ref{fig:omegac}), which increases 
exponentially  above the characteristic force $F_0$.

In the simulations of MT ensembles under force in 
Ref.\ \cite{Laan2008}, this catastrophe model is used 
 in  combination with 
a  characteristic force $F_0  \simeq 0.8\;{\rm pN}$
based on the experimental data in Ref.\
\cite{DY97}; this value 
 is significantly smaller  than  the theoretical value $F_0 = k_BT/d
\simeq 7\;{\rm pN}$.
For simulations,  a 10-fold increased catastrophe rate is used in 
Ref.\ \cite{Laan2008}. 
The other parameter values  used in  Ref.\ \cite{Laan2008}
are $v_{\text{on}} = \omega _{\text{on}}d = 4.37\times 10^{-8}\;{\rm m/s}$, 
 which corresponds to $\omega_{\text{on}}\approx 70\;\rm{s}^{-1}$
\cite{Laan2008,DY97}. 
We will use the same set of parameters, apart from the 10-fold increase 
of the catastrophe rate, i.e., 
a small value $F_0 \simeq 0.8\;{\rm pN}$  for our simulations with 
the catastrophe model by Janson {\it et al.}

Both  the Flyvbjerg and Janson catastrophe models  describe available 
experimental data on {\it single}  MTs, as has been shown in Refs.\ 
\cite{Flyv1994PRL,Flyv1996PRE} and \cite{Janson2003}, respectively. 
Therefore, it is important to investigate whether they also 
give compatible results for the {\it collective} dynamics of MTs 
under force.

In both models the catastrophe rate $\omega_c$ 
decreases as a power law over a wide range of growth velocities,
\begin{equation}
   \omega_c \propto v_+^{-2/3}~~ (\mbox{Flyvbjerg}),~~~~~~ 
    \omega_c \propto      v_+^{-1}  ~~  (\mbox{Janson})
 \label{omegacv+}
\end{equation}
Because of $v_+ \sim d\omega_{\rm on} e^{-F/F_0}$ for large velocities,
$\omega_c$  increases exponentially with force $F$ 
above the characteristic force $F_0$ 
in both catastrophe models. 
This can also be seen in the comparison in Fig.\ \ref{fig:omegac}.

Our theory will predict 
 collective catastrophe and rescue oscillations of the MT ensemble 
for all catastrophe models which fulfill two conditions:
\begin{itemize}
\item[(i)] Force dependence via growth velocity:
 The catastrophe rate 
$\omega_c=\omega_c(v_+)$ is a function of the growth velocity and 
 becomes force-dependent 
via the force dependence of the growth velocity, $\omega_c = \omega_c(v_+(F))$;
if single-MT catastrophes are related to hydrolysis within the MT and 
the loss of the stabilizing GTP cap, $\omega_c$ 
should be a {\em decreasing} function of $v_+$ and, thus, 
an {\em increasing} function of the force $F$. 
\item[(ii)] Exponential force dependence above $F_0$:
  The resulting force dependence of $\omega_c$ is such that 
$F d\omega_c/dF \gg \omega_c(F)$ for  $F>F_0$  above the characteristic 
force $F_0$ [see also Eq.\ (\ref{criticalFc})].
This gives rise to a catastrophe rate that increases  
exponentially   with force above the characteristic force $F_0$.
\end{itemize}
Requirement (ii) is fulfilled for all catastrophe rates
decreasing as a power-law  $\omega_c \propto v_+^{-\varepsilon}$ 
($\varepsilon>0$) with 
growth velocity as in both the 
Flyvbjerg and  Janson models; see Eq.\ (\ref{omegacv+}). 
Accordingly,  we will find collective catastrophe and rescue 
oscillations for both models.

It is possible to consider other types of 
catastrophe models where the catastrophe rate $\omega_c=\omega_c(v_+)$ 
is a decreasing function of the growth velocity according to requirement
(i) but where requirement (ii) of an exponentially increase 
of the catastrophe rate with force is violated. 
One particularly simple example  is 
a catastrophe rate which  decreases {\it linearly} with velocity,
\begin{equation}
  \omega_c(v_+) = \tilde{a} - \tilde{b} v_+.
\label{linear}
\end{equation}

We will demonstrate that collective catastrophe and rescue 
oscillations are indeed absent for such a ``linear model.''

We will not discuss more elaborate 
 multistep catastrophe models with more than two MT states,
which have been proposed only recently 
 \cite{Gardner2011}.

\subsection{Model for  MT ensemble}

\begin{figure}
\begin{center}
  \epsfig{file=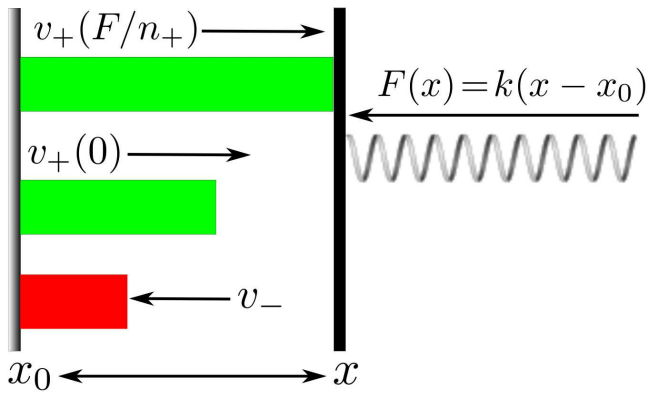,width=0.48\textwidth}
\caption{\label{fig:model}
(Color online) Schematic representation of $N=3$ MTs for 
   $n_+=1$. From top to bottom: 
  MTs  under force $F=k(x-x_{0})$ grow with velocity
  $v_{+}({F}/{n_{+}})$ and
   unloaded MTs with  $v_{+}(0)$.  
  MTs shrink with $v_{-}$ after  a catastrophe. 
}
\end{center}
\end{figure}

We consider an ensemble of $N$ parallel MTs, 
directed along the $x$-direction. 
The  ensemble is growing in a positive $x$-direction and 
 pushing against an elastic  barrier, as shown 
 in  Fig.\ \ref{fig:model}.
The cooperative dynamics is governed by 
the number $n_{+}<N$ of leading MTs
which push  simultaneously in the growing state. 

The elastic barrier is modelled as a spring 
 with  equilibrium position 
$x_0= 1\;{\rm \mu m}$ 
and a  spring constant $k$ in the range 
$10^{-7}\;{\rm N/m}$ (soft) to $10^{-5}\;{\rm N/m}$ 
(stiff as in the optical trap 
experiments in Ref.\ \cite{Laan2008}).
Barrier displacement 
 by  the leading MTs with their tips positioned at $x>x_{0}$ 
causes a force  $F=F(x)=k(x-x_{0})$ resisting further growth;
for $x<x_0$ there is a force-free region. 
We assume that the force $F$ is equally 
 shared between all $n_{+}$ leading MTs
such that each leading MT is subject to a force $F/n_{+}$.
Force-sharing is the only coupling between the MTs. 
In the presence of rescue events, i.e., for nonzero rescue rate, 
we  force MTs shrinking to $x=0$ to undergo rescue.

Under a shared force $F/n_+$, 
the growth velocity of a MT  reduces to 
$v_{+}({F}/{n_{+}})=d \left[ \omega_{\text{on}}
   e^{-F/n_{+}F_0}-\omega_{\text{off}} \right]$
with the  characteristic force  $F_0=k_BT/d$  
governing monomer attachment; see Eq.\ (\ref{v+F}).
The force-dependent growth velocity also 
  gives rise  to a  catastrophe rate
$\omega_c=\omega_c(v_+(F/n_{+}))$ increasing with force.
All nonleading MTs grow with the higher zero force velocity
$v_{+}(0)$  in their growing state. 
Therefore,  nonleading MTs, which grow 
force free and fast,  ``catch up'' 
leading MTs, which grow under force with reduced velocity. 
This mechanism supports a state of collective growth,
where a relatively large number $n_+$ of MTs are pushing 
cooperatively.  
We  assume that the shrinking velocity 
$v_{-}$ is independent of  force.
Because the catastrophe rate depends on force only via $v_+$, the relevant 
force scale of the problem is set by the characteristic
 force $F_0$; see (\ref{v+F}).

This model for the dynamics of the MT ensemble is 
 very similar  to the model 
underlying the simulations in Ref.\ \cite{Laan2008}.
In particular, we use the same rules for the coupling between
MTs by the load force. 
The most important difference is that we include rescue events in 
the single MT dynamics, which have not been considered in 
Ref.\ \cite{Laan2008}.

\section{Model parameters and simulation}
\label{sec_parameters}

In the simulation model the following parameters for MT
growth have to be specified:

(i)  The  effective GTP-tubulin monomer size $d$, which is 
$d= 8\,{\rm nm}/13 = 0.6\;{\rm nm}$
  for 13 protofilaments.

(ii) The characteristic force $F_0$ 
  governing the exponential decrease (\ref{v+F}) of the single 
  MT growth velocity with force. 
   We use $F_0 =  k_BT/d \simeq 7\;{\rm pN}$ in the Flyvbjerg 
  catastrophe model  and the measured value 
  $F_0 \simeq 0.8\;{\rm pN}$ \cite{DY97} in the Janson catastrophe 
  model. 

(iii) We use a  GTP-tubulin monomer 
  off-rate of $\omega_{\text{off}}= 6\;{\rm s}^{-1}$ 
  \cite{Janson2004}.
 Then  the growth velocity in the  absence of force 
    $v_+(0)=d(\omega_{\text{on}}-\omega_{\text{off}})$
  determines 
    the GTP-tubulin monomer  on-rate 
  $\omega _{\text{on}}$ (which is  proportional to the 
  GTP-tubulin concentration $C_T$).

(iv) 
  Within the cooperative hydrolysis model
   by Flyvbjerg  \cite{Flyv1994PRL,Flyv1996PRE}, 
    the catastrophe rate $\omega_c$ is determined  by the 
 growth velocity  
$v_{+}$ [according to (\ref{v+F})  with $F_0 =  k_BT/d \simeq 7\;{\rm pN}$]
   and the hydrolysis parameters 
 $v_h = 4.2\times 10^{-9}\;{\rm m/s}$ and 
$r = 3.7\times 10^{6}\;{\rm m}^{-1}{\rm s}^{-1}$
 \cite{Flyv1996PRE}. 
The exact catastrophe rate  is calculated from the numerical 
 solution of Eq.\
(\ref{Ai}) for given $v_+$, $v_h$, and $r$, as explained 
in Sec. \ref{sec_catmodels} above.

(v) 
  Within the hydrolysis model
   by Janson \cite{Janson2003}, 
    the catastrophe rate $\omega_c$ is determined  by the 
   growth velocity  
   $v_{+}$ [according to (\ref{v+F})  with $F_0 \simeq 0.8\;{\rm pN}$] 
  and the  parameters  $a\simeq 20\;{\rm s}$ and 
   $b\simeq 1.4\times 10^{10}\;{\rm s}^2m^{-1}$.

(vi) The shrinking velocity  $v_- = 3\times 10^{-7}\,{\rm m/s}$.

(vii) The rescue rate $\omega_r$.

(viii) 
  For the elastic force $F=k(x-x_0)$ on the leading MTs we use a
    spring stiffness $k= 10^{-7}\;{\rm  N/m}$ for a 
  soft elastic barrier and  $k= 10^{-5}\;{\rm  N/m}$ for 
  a stiff elastic barrier as in the optical trap 
experiments in \cite{Laan2008}. The rest 
  position $x_0$ of the spring is taken as $x_0=1\;{\rm \mu m}$.
  For nonzero rescue rate, we also 
 use reflecting boundary conditions at $x=0$, i.e., 
   a filament undergoes immediate rescue if it shrinks to $x=0$.

The parameters $v_-$,  $\omega_{\text{off}}$, $v_h$, and $r$ are 
fixed in simulations. 
We vary the on-rate $\omega_{\text{on}}$ and, thus, 
$v_+(0)$  and the rescue rate $\omega_r$ within parameter
ranges,
which are selected according to literature 
values collected in Table \ref{tab:parameters}.

\begin{table*}
\begin{center}
 
 \begin{tabular}{|c|c|c|c|c|}
\hline
\hline
Ref.\ & $v_{+}(0)$ [m/s] & $\omega_{\text{on}}$ [1/s] &  $v_{-}$ [m/s] & $\omega_{r}$ [1/s]\\
\hline

Drechsel \cite{Drechsel1992} & $(0.7\;...\;2)\times 10^{-8}$ & $(11\;...\;32)$ & $\sim 1.8\times 10^{-7}$ & - \\

Gildersleeve \cite{Gildersleeve1992} & $\sim 4.2\times 10^{-8}$ & $\sim 68$ & $\sim 4.2\times 10^{-7}$ & -\\

Walker \cite{Walker1988} & $ (4\;...\;8)\times 10^{-8}$ & $(63\;...\;130)$ & $\sim 5\times 10^{-7}$ & $(0.05\;...\;0.08) $(TUB)\\

Laan \cite{Laan2008} & $\sim 4.2\times 10^{-8}$ & $68.25$ & - & - \\ 

Janson \cite{Janson2004} & $(3\;...\;4.3)\times 10^{-8}$ & $(53\;...\;74)$ & - & - \\

Pryer \cite{Pryer1992} & - & - & - & $...\;0.5$ (TUB)  $..\,0.15$ (MAPS) \\

Dhamodharan \cite{Dhamodharan1995} & - & - & - & $ ...\;0.07$ (Cell) $\quad...\;0.085$ (MAPS) \\

Nakao \cite{Nakao2004} & - & - & - & $...\;0.1$ (TUB) \\

Shelden \cite{Shelden1993} & - & - & - & $(0.03\;...\;0.2)$ (Cell) \\
\hline
\hline
\end{tabular}
\caption{Literature values for MT growth 
parameters $v_{+}(0)$,  $\omega_{\text{on}}$,
  $v_{-}$, and  $\omega_{r}$.
  TUB: {\it in vitro} results for tubulin solutions, Cell: {\it in vivo}
  results,  MAPS: effect from MT associated proteins. 
  Values for $\omega_{\text{on}}$
 are estimated from measured growth velocities via $\omega_{\text{on}} \approx
 v_{+}(0)N/d$ neglecting $\omega_{\text{off}}$.}
\label{tab:parameters}
\end{center}

\end{table*}

In the simulation, we integrate the deterministic equation of motion 
for an ensemble of $N$ MTs with continuous lengths $x_i$ ($i=1,\ldots,N$)
and include stochastic 
switching between growth and shrinking for each MT.
In the integration we use a fixed time step $\Delta t = 0.1\;\rm{s}$,
which is small enough to ensure $\omega_{c,r}\Delta t\ll 1$.
 In each time step, we have to determine the number 
$n_+$ of leading force-sharing MTs. 
This is done by regarding all growing MTs within a 
distance $v_+(F/n_+)\Delta t$ of the leading 
MTs as leading for the next time step. 
We can perform two kinds of averages:  
Averages $\langle \ldots \rangle$ are taken over many 
realizations, and averages $\overline{...}$ are timeaverages.

\section{Collective catastrophes at zero rescue rate}
\label{no_rescue}

We start the analysis with the case of 
zero rescue rate because this case is  conceptually 
simpler to understand as rescue events are absent, and 
there are only collective catastrophes to be discussed.
Furthermore, this case is particularly important because 
experimental data are available:
In recent experiments, Laan {\it et al.}\ \cite{Laan2008} 
showed that MT ensembles 
exhibit  phases of collective growth followed by 
 collective catastrophes, where all leading MT 
nearly simultaneously undergo a catastrophe.  
The experiments were  performed 
on short time scales such that no rescue events occur.  
 It was also observed  that the maximal polymerization  
 force before catastrophes  grows linearly in $N$. 
We quantify these 
features based on  a dynamical mean-field theory.

In an ensemble of $N$ MTs 
the dynamic instability of individual MTs leads to
 stochastic fluctuations in the number $n_{+}$ 
of leading MTs. The force $F$ changes by filament growth 
according to $\dot{F} = k \dot{x}$ with $\dot{x}=v_+(F/n_+)$ 
if the ensemble grows ($n_+\ge 1$) 
and $\dot{x}=-v_-$ if all MTs shrink ($n_+=0$).
 In a state of collective growth,
 a stable mean number  of  
MTs are pushing cooperatively, while the force $F$ is 
increasing by  growth against the elastic barrier.
If the number $n_{+}$ of pushing MTs 
is reduced by an individual catastrophe, the force on the 
remaining $n_+-1$ leading MTs increases and, thus,  their
catastrophe rate $\omega_c(F/n_+)$ increases. 
A cascade of individual 
catastrophes \textemdash a {\em collective catastrophe} \textemdash  can be initiated 
until a state  $n_+=0$ is reached with all MTs shrinking.
This is the final absorbing state of the system in the absence 
of rescue events.

The stochastic
dynamics of $n_{+}$ in a growing phase in the absence of rescue events
is described by a
one-step master equation with backward rates 
 $r_{n_+}=n_+\omega_{c}(F/n_{+})$ for  decreasing $n_+$ by one,
which derive from the catastrophe rate of individual MTs under
force sharing. 
In a mean-field approach, we replace the stochastic 
variables $F$ and 
$n_+$ by their (time-dependent)
 mean values $\langle F \rangle$ and $\langle n_{+}\rangle$
(averaging over many realizations of the stochastic $n_+$-dynamics)
and neglect all higher-order correlations, e.g.;\ set 
$\langle F/n_+ \rangle = \langle F \rangle/\langle n_+ \rangle$. 
In the growing phase, 
 we then obtain  two coupled mean-field equations,
\begin{eqnarray}
  d{\langle n_{+}\rangle}/dt&=&
 -\langle n_{+} \rangle
   \omega_c\left(\langle F\rangle/\langle n_{+}\rangle \right),
 \label{eq:avgn1}\\
  d{\langle F\rangle}/dt &=& k 
     v_+\left(\langle F\rangle/\langle n_{+}\rangle \right).
\label{eq:avgF}
\end{eqnarray}
In the mean-field approximation
we can calculate the maximal polymerization force $F_{\rm max}$
(averaged over many realizations)
that is reached 
during  the mean first passage time  from $n_+=N$ to 
$n_+=0$  by solving 
\begin{equation}
 \frac{d\langle F \rangle}{d\langle n_+ \rangle} = 
  \frac{d\langle F\rangle/dt}{d\langle n_+\rangle/dt} = 
  -\frac{k v_+\left(\langle F\rangle/\langle n_{+}\rangle \right)}
   {\langle n_{+} \rangle 
  \omega_c\left(\langle F\rangle/\langle n_{+}\rangle \right)}.
\end{equation}
with initial conditions 
$\langle F\rangle=0$ for $\langle n_+ \rangle=N$ in order 
to find $\langle F\rangle=F_{\rm max}$  at $\langle n_+ \rangle\approx 0$.

Above the characteristic force $F_0$, the ratio
$v_+(F)/\omega_c(F)$ decays exponentially because $v_+(F)$ decreases 
exponentially and $\omega_c(v_+(F))$ increases  exponentially.
Therefore,  we can solve 
 in two steps:
(i) As long as the shared force is small compared to $F_0$, 
$\langle F\rangle/\langle n_+ \rangle \ll F_0$, 
we neglect the force and find 
$\langle F\rangle \approx  kv_+(0)/\omega_c(0) \ln(N/\langle n_+ \rangle)$.  
(ii)  For $\langle F\rangle/\langle n_+ \rangle \gg F_0$,  
on the other hand, the catastrophe frequency 
increases exponentially, and we can assume that 
$d\langle F \rangle/d\langle n_+ \rangle \approx 0$ and 
$\langle F \rangle$ remains constant.

\begin{figure}
\begin{center}
  \epsfig{file=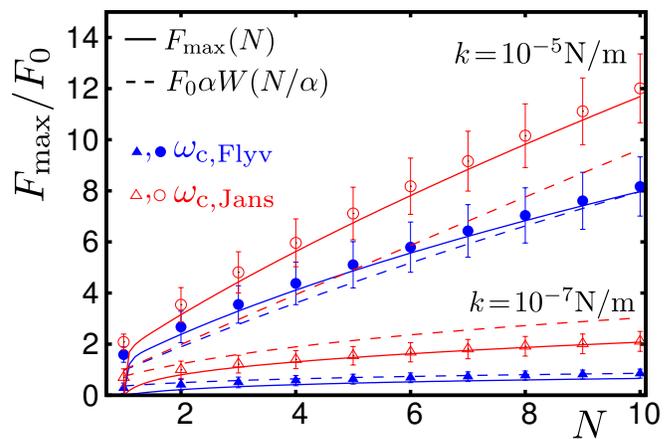,width= 0.48\textwidth  }
\caption{\label{fig:force_norescue}
(Color online)
 $F_{\text{max}}/F_0$ as a function of $N$ for zero rescue rate, 
   $\omega_{\text{on}}=70\;{\rm s}^{-1}$, 
    and 
 $k=10^{-7}\;{\rm N/m}$ ($\alpha\simeq 2.7$, soft barrier)
 and $k=10^{-7}\;{\rm N/m}$ ($\alpha\simeq 27$, stiff barrier).
We compare results for the catastrophe models of Flyvbjerg (blue, solid symbols) 
and Janson (red, open symbols). 
Data points are simulation results; 
error bars represent the  standard deviation of the 
stochastic quantity $F_{\text{max}}/F_0$.  
Solid lines are numerical solution
of the mean field dynamics (\ref{eq:avgn1}) and (\ref{eq:avgF}).
Dashed lines are analytical estimates 
according to (\ref{Fmaxzero}).
}
\end{center}
\end{figure}

The boundary between regimes (i) and (ii) 
is determined by the condition 
  $\langle F\rangle/\langle n_+ \rangle=F_0$:
 Regime (i) applies for
$\langle n_+ \rangle > n_0$ with 
$ n_0 = \alpha W(N/\alpha)$,
 where 
$W(x)$ is the Lambert $W$-function, which  is the solution of 
$x = We^W$ (for $W \ge -1$). 
The parameter  
\begin{equation}
\alpha \equiv kv_+(0)/\omega_c(0)F_0
\label{alpha}
\end{equation}
is a dimensionless measure for  the stiffness of the elastic barrier.
Because $\langle F \rangle$ remains constant 
for $\langle n_+ \rangle < n_0$,
the resulting maximal polymerization force is given by 
\begin{equation}
    F_{\rm max} = n_0 F_0 = F_0 \alpha W(N/\alpha),
\label{Fmaxzero}
\end{equation}
with a logarithmic 
 asymptotics $F_{\rm max}  \approx F_0\alpha \ln(N/\alpha)$ 
for large  $N\gg \alpha$ or a soft barrier and a quasilinear behavior 
$F_{\rm max}  \approx F_0 N (1- N/\alpha)$, which is {\em independent}
of $\alpha$ to leading order,
for small $N\ll \alpha$ or a stiff barrier.
The mean-field result (\ref{Fmaxzero}) agrees with  
numerical solutions of the mean field dynamics 
as given by Eqs.\ (\ref{eq:avgn1}) and (\ref{eq:avgF}) and 
full stochastic simulations
both for soft and stiff barriers, as can be 
 seen in  Fig.\ \ref{fig:force_norescue}.

The parameter $\alpha$ can also be interpreted as a measure for the 
relative speed of the  initial $\langle n_+ \rangle$ and
 $\langle F\rangle$ dynamics according to the mean-field 
equations\ (\ref{eq:avgn1}) and (\ref{eq:avgF}),
which allows us to give simple arguments for the 
maximal polymerization force $F_{\rm max}$: 
For $\alpha \ll N$ (the case of a soft barrier), 
the $\langle n_+\rangle$ dynamics is fast compared to the
 $\langle F\rangle$ dynamics.
Therefore, $\langle n_+\rangle$ decays approximately force free 
in a time $t_c \sim 1/\omega_c(0)\ln N$ from $\langle n_+\rangle=N$ 
to $\langle n_+\rangle=1$. During this time, the force reaches
a value $F_{\rm max} \sim kv_+(0) t_c \sim F_0 \alpha \ln N$.
For  $\alpha \gg N$ (the case of a stiff barrier), 
the $\langle n_+\rangle$-dynamics is initially slow compared to the
 $\langle F\rangle$-dynamics and $\langle n_+\rangle\approx N$ until
the characteristic force $F_0$ per MT is reached and the catastrophe rate 
increases exponentially. Up to this point, essentially 
 $N$ MTs share the force such 
that $F$ increases up to $F_{\rm max} \sim F_0N$ until catastrophes  
set in. This takes a time $t_c \sim NF_0/kv_+(0)$ and 
$\Delta\langle n_+\rangle \sim  \dot{\langle n_+\rangle}t_c \sim 
  N/\alpha \ll 1$ is indeed small such that the  assumption 
$\langle n_+\rangle\approx N$  is consistent.

In the experiments in Ref.\ \cite{Laan2008}, the spring stiffness
was  $k\simeq 10^{-5}\;\rm{N/m}$, which gives $\alpha\simeq 27$ such that 
 these experiments were performed  in the 
quasilinear regime of a stiff barrier, where we 
predict $F_{\rm max}  \approx F_0 N$
for all experimentally accessible $N$ 
(see the upper lines in Fig.\ \ref{fig:force_norescue}).
This linear increase is in agreement with the experimental results
but the ratio $F_{\rm max}/N$ is only of the order of $3\;{\rm pN}$ 
experimentally, while $F_0 \simeq 7\;{\rm pN}$. This 
hints at a lower value for $F_0$ in the force-polymerization velocity 
relation for MTs; experimentally, a value $F_0 \simeq 2\;{\rm pN}$ has been 
measured in Ref.\ \cite{DY97}, which is indeed compatible with 
the experimental results of  Ref.\ \cite{Laan2008}.

In Fig.\ \ref{fig:force_norescue}, we  compare
 mean field theory and simulation results  for the Flyvbjerg and the Janson 
catastrophe model for both soft and stiff barriers. 
For both models, we find agreement
between mean-field theory and simulations and, moreover, 
 both models give comparable values for generated forces. 
This demonstrates that results for the cooperative force generation at 
zero rescue rate are robust with respect to  details of the 
single-MT catastrophe model.  The essential feature 
entering the mean-field theory is the exponential increase of the 
catastrophe frequency  with force above the characteristic force $F_0$.

\section{Collective catastrophes and collective rescue 
 for nonzero rescue rate}
\label{sec_rescue}

We now consider force generation in the presence of rescue events. 
Rescue events were not included in the simulation model 
in Ref.\ \cite{Laan2008}.  Also experiments in Ref.\ \cite{Laan2008}
were  performed 
on short time scales such that no rescue events occurred. 
However, rescue events are an essential part of MT dynamics, 
 and their influence on force generation 
and MT dynamics needs to be addressed.

In the presence of rescue events, 
the dynamics will not change considerably for $N\ll \alpha$,
i.e., for a stiff barrier because this limit corresponds to a {\em slow}
$\langle n_+ \rangle$ dynamics, which cannot benefit from 
additional rescue events. Moreover, because the 
$\langle F \rangle$ dynamics is fast, 
rescue will not  happen before  
the force-free region $x<x_0$ is reached, where MTs decouple.

Therefore, we focus on the influence of 
rescue events for $N \gg \alpha$ or a soft barrier
corresponding to a {\em fast} $\langle n_+ \rangle$ dynamics.
In this regime,  the collective dynamics becomes
strongly modified. Apart from collective catastrophes 
also collective rescue events occur:
After a collective catastrophe the system is in a state 
 $n_+=0$ with all MTs shrinking.  Individual rescue 
 events lead to $n_+=1$, but a single MT bearing the whole force undergoes
  an immediate catastrophe again
 with high probability. Therefore, a 
 cascade of rescue events \textemdash a  {\em collective rescue} \textemdash 
 is necessary to increase $n_+$ back to  
  a number sufficient to maintain stable
  collective growth.

\subsection{Simulation results}

\begin{figure}
 \begin{center}
 \epsfig{file=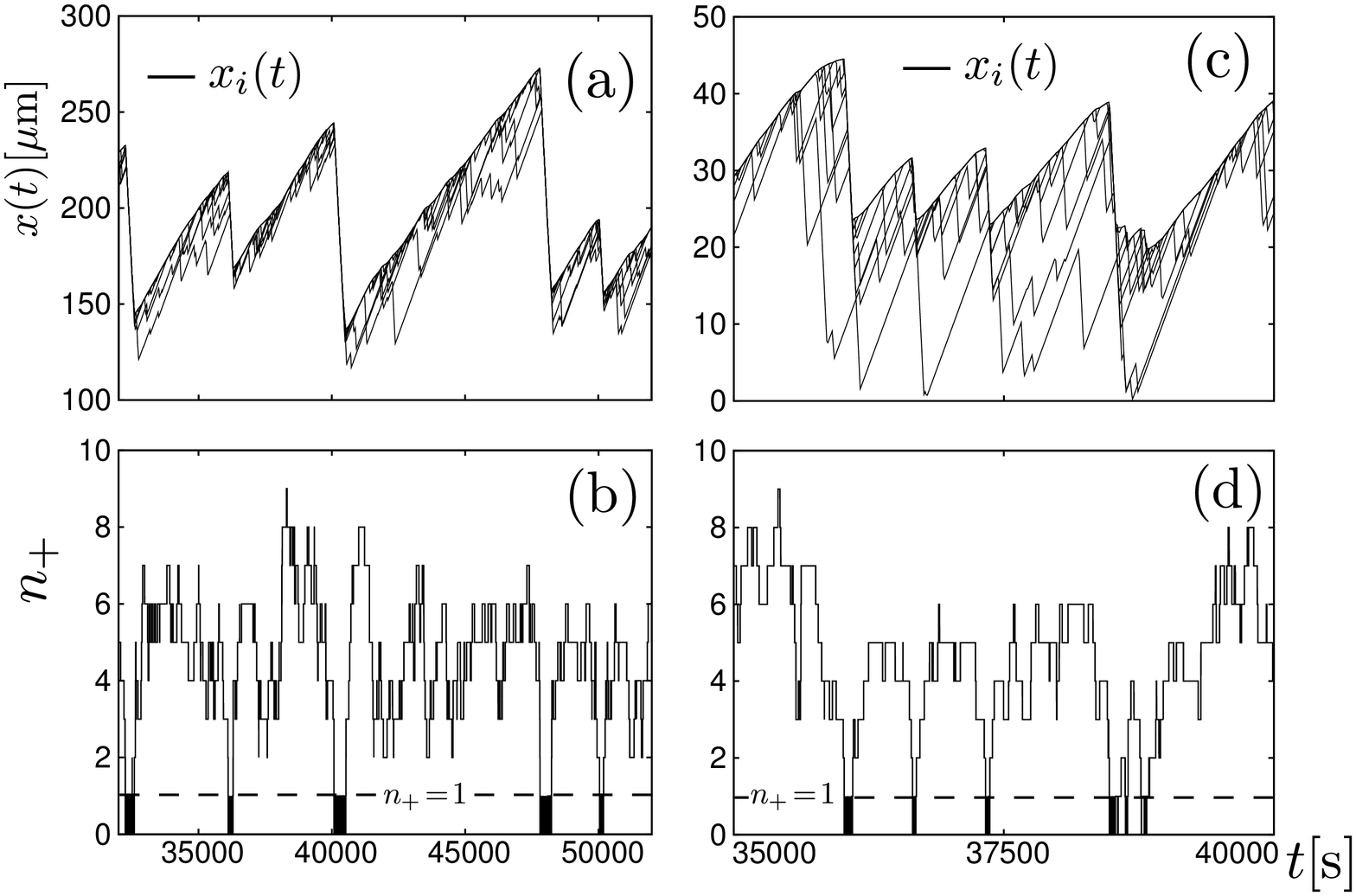,width=0.45\textwidth}
  \caption{\label{fig:trajectories} 
 Typical simulation trajectories for $N=10$ MTs, 
   $\omega_{\text{on}}=70\;\rm{s}^{-1}$, 
 and $\omega_{\text{r}}=0.05\;\rm{s}^{-1}$; left  [(a) and (b)]
  for the Flyvbjerg catastrophe model and right [(c) and (d)] for the 
  Janson catastrophe model. 
 [(a) and (c)] Positions  of all MTs as a function of time $t$; the obstacle 
  position $x(t)$ is the position of the leading MT. 
  [(b) and (d)] The number $n_+$ 
  of leading MTs as a function of time $t$. 
Collective catastrophes and collective rescue events  can be clearly 
recognized:
In a collective catastrophe $n_+$ drops to $n_+=1$ 
   and $x(t)$ of the leading MTs 
  starts to shrink; after a collective rescue $n_+$  starts to 
 increase again to values $n_+>1$, and $x(t)$ of the leading MTs 
  start to grow. 
}
 \end{center}
\end{figure}

Alternating collective catastrophes and collective rescue 
events give rise to oscillations in the polymerization force 
or, equivalently, the position $x$ of the obstacle.
Such oscillations with alternating 
 collective catastrophes and collective rescue events 
can be clearly seen in the stochastic simulation trajectories 
for the positions $x$ of the  MTs [see Figs.\ \ref{fig:trajectories}(a) and\ \ref{fig:trajectories}(c)]
and the number $n_+$ 
  of leading MTs [see Figs.\ \ref{fig:trajectories}(b) and\ \ref{fig:trajectories}(d)]
 as a function of time $t$. The simulation trajectories 
also show that this phenomenon is robust with respect to the 
catastrophe model and can be observed both for the Flyvbjerg and the 
Janson catastrophe models, which are shown in Fig.\ \ref{fig:trajectories}
on the left and the right sides, respectively, and exhibit qualitatively
very similar behavior. 
Similar oscillations have been observed in the simulations 
in Ref.\ \cite{Laan2008} in the presence of 
 MT renucleation instead of MT rescue  and for a 
constant force.

\begin{figure*}
 \begin{center}
   \epsfig{file=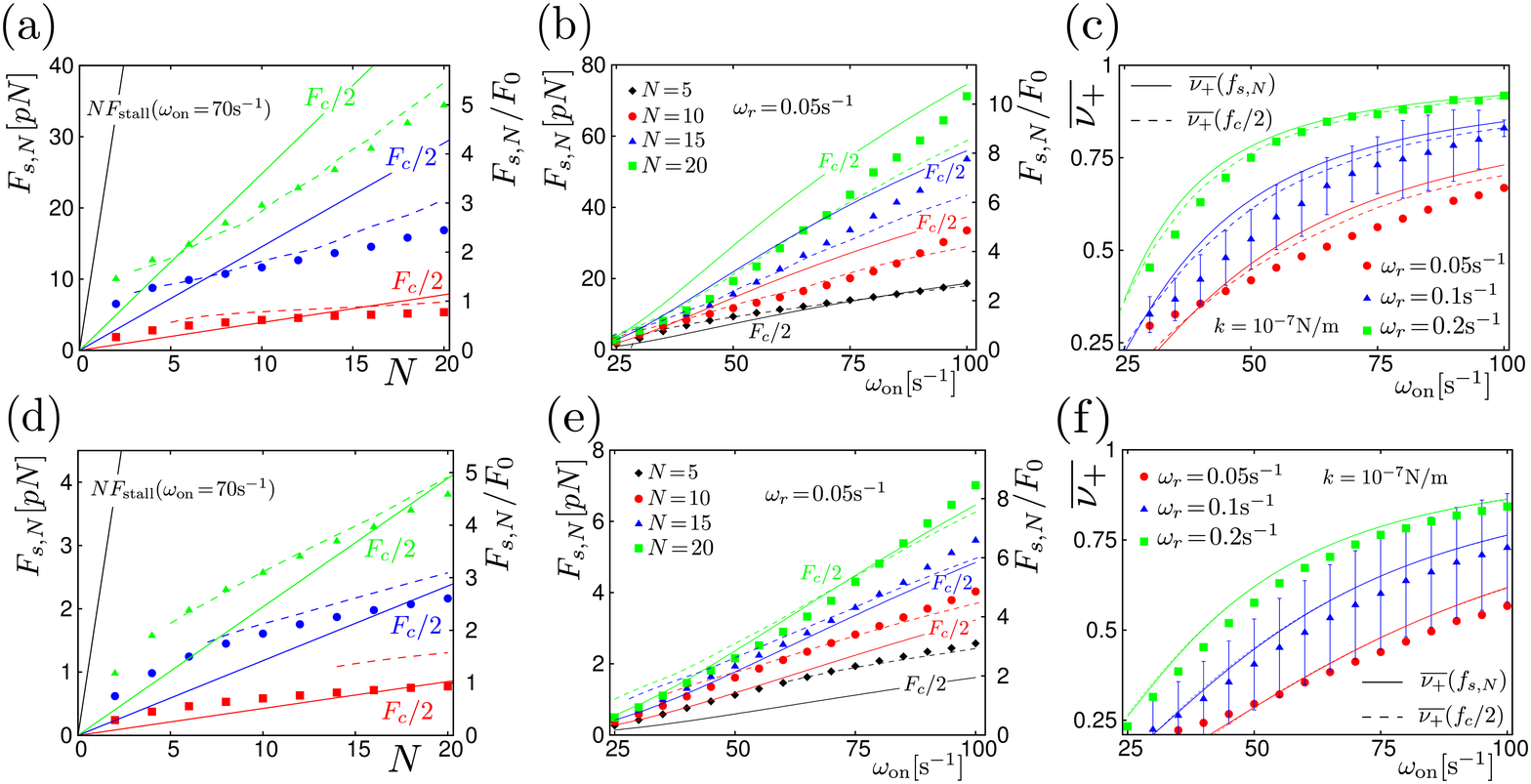,width=\textwidth}
 \caption{\label{fig:force}
(Color online)
 Upper row: Simulation results using the Flyvbjerg catastrophe model. 
    Lower row:   Simulation results using the Janson  catastrophe model.
  Both are  for a soft barrier ($k=10^{-7}\;{\rm N/m}$).
  [(a) and (d)]  Simulation results for the average polymerization force 
  $F_{\text{s,N}}$ as a function of the number $N$ of MTs 
   for different on-rates
   $\omega_{\text{on}}= 30\;{\rm s}^{-1}\; (\blacksquare)$, 
  $50\;{\rm s}^{-1}\; (\bullet)$,
   and 
   $70\;{\rm s}^{-1}\; (\blacktriangle)$ at a   fixed rescue rate 
   $\omega_{r}=0.05\;{\rm  s}^{-1}$. 
   Solid lines: Mean field estimate $F_{s,N}=F_{c}/2$, see Eq.\
   (\ref{avF}) 
  (neglecting $F_{\text{min}}$). 
   Dashed lines: Numerical mean field solution including stochastic effects. 
   Black solid line:  $N$-fold single-MT stall force $F_{s,N}= N F_{\rm stall}$
   for $\omega_{\text{on}}= 70\;{\rm s}^{-1}$.
 [(b) and (e)] Polymerization force $F_{\text{s,N}}$ as a
   function of  on-rate $\omega_{\text{on}}$ for
     $N=5\; (\blacklozenge), 10\; (\bullet), 15\; (\blacktriangle), 
     20\; (\blacksquare)$, and $\omega_{r}=0.05\;{\rm s}^{-1}$.
   Solid lines: Mean field estimate $F_{s,N}=F_{c}(\omega_{\text{on}})/2$.
   Dashed lines: Numerical mean-field solution including stochastic effects. 
 [(c) and (f)]  Time-averaged pushing fraction
   $\overline{\nu_{+}}= \overline{\langle n_+ \rangle}/N$ 
    of MTs as a function of $\omega_{\text{on}}$ for
   $\omega_{r}=0.05\;{\rm s}^{-1}\; (\bullet)$, $0.1\;{\rm s}^{-1}\;
   (\blacktriangle)$, and $0.2\;{\rm s}^{-1}\; (\blacksquare)$. 
   Solid lines: Solution of Eq.\ (\ref{eq:steady_nu}) for $f=f_{c}/2$.
   Dashed lines: Solution of Eq.\ (\ref{eq:steady_nu})
       for $f=F_{\text{s,N}}/N$ with $F_{\text{s,N}}$ from the  
    numerical mean-field solution including stochastic effects.  
    Error bars represent the standard deviation  of the stochastic quantity
   $\overline{n_{+}}/N$. For reasons of clarity we only 
    show error bars for $\omega_{r}=0.1\;\rm{s}^{-1}$. 
   All other standard deviations are of the same magnitude.
}
  \end{center}
 \end{figure*}

In the simulations we measure the polymerization 
force $F_{\text{s,N}} = \overline{\langle F \rangle}$, which is averaged
over time and many realizations as a 
function of MT number $N$ and of the on-rate $\omega_{\text{on}}$.
This  time-averaged polymerization 
force is  also the stall force of the MT ensemble. 
The results are shown in Fig.\ \ref{fig:force} 
for both the Flyvbjerg catastrophe model [upper row: Figs.\ \ref{fig:force}(a) and\ \ref{fig:force}(b)] and
 the Janson catastrophe 
model [lower row: Figs.\ \ref{fig:force}(d) and\ \ref{fig:force}(e)]. 
The main finding of the simulations is an approximately {\em linear}
increase of the polymerization 
force $F_{\text{s,N}}$ with  the number $N$ of MTs
 [see Figs.\ \ref{fig:force}(a) and\  \ref{fig:force}(d)].
This shows that for large MT ensembles,
 rescue events give rise 
 to much higher polymerization 
forces  as compared to the  logarithmic 
$N$ dependence derived in the previous 
section  in the absence 
of rescue events for a soft barrier  ($N\gg \alpha$).
Simulations also show an approximately linear 
  increase of the 
polymerization force with the on-rate $\omega_{\text{on}}$
 [see Figs.\ \ref{fig:force}(b) and\ \ref{fig:force}(e)].

We also show numerical results for the time-averaged 
 pushing fraction
   $\overline{\nu_{+}}= \overline{\langle n_+ \rangle}/N$ 
    of MTs as a function  of the on-rate $\omega_{\text{on}}$
in Figs.\ \ref{fig:force}(c) and \ref{fig:force}(f).
The pushing fraction increases with on-rate, which demonstrates
an increasing tendency of MTs to push {\it synchronously} at 
higher on-rates, where larger forces are generated. 

 Simulation results for the  Flyvbjerg model 
(upper row in Fig.\ \ref{fig:force})  and the Janson model 
(lower row in Fig.\ \ref{fig:force})
 show a very similar linear increase 
 for the polymerization force $F_{\text{s,N}}$ with $N$ 
and very similar results for the  time-averaged  pushing fraction
   $\overline{\nu_{+}}$  of MTs, which is in accordance with the 
qualitatively similar  simulation trajectories shown in 
 Fig.\ \ref{fig:trajectories} for both catastrophe models. 
This further supports that our results are  robust with respect to the 
catastrophe model. 
The absolute values of typical forces in Figs.\ \ref{fig:force}(a),\ \ref{fig:force}(b),\ \ref{fig:force}(d), and\ \ref{fig:force}(e) 
and, similarly, between typical MT lengths in 
Figs.\ \ref{fig:trajectories}(a) and\ \ref{fig:trajectories}(c) differ, however,
between the two catastrophe models. The reason is that the basic force 
scale of the problem is the characteristic force $F_0$, above which 
the catastrophe rate increases exponentially, as will be shown 
below.  
We have chosen the theoretical value $F_0 = k_BT/d \simeq 7\;{\rm pN}$
for the Flyvbjerg model and the much smaller value 
$F_0 = 0.8\;{\rm pN}$ according to  Ref.\ \cite{Laan2008} 
with the Janson model.
In units of the  characteristic force $F_0$,  typical  forces
are very similar [see  Figs.\ \ref{fig:force}(a),\ \ref{fig:force}(b),\ \ref{fig:force}(d), and \ \ref{fig:force}(e) right scale].

\subsection{Dynamical mean field theory}

We will show that all  simulation results and the robustness 
with respect to the catastrophe model can be explained 
based on a dynamical mean-field theory. 

\begin{figure}
\begin{center}
  \epsfig{file=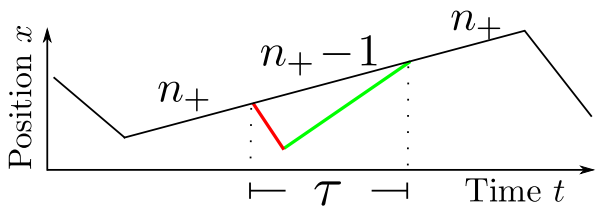,width=0.9\columnwidth}
\caption{\label{fig:tau}
(Color online)
Illustration of time scale $\tau$ from Eq.\ (\ref{eq:tau}).
}
\end{center}
\end{figure}

In the presence of rescue events, 
the mean-field equation (\ref{eq:avgn1}) for 
$\langle n_{+}\rangle$ becomes modified in the growing phase.
The one-step master equation for  
 $n_{+}$ in a growing phase also contains a  forward rate 
 $g_{n_+}=(N-n_{+})\tau^{-1}$  for increasing $n_+$ by one. 
 This forward rate 
 is determined by a rescue and ``catch-up''process for 
 the $(N-n_+)$ MTs, which are not pushing:
 The time  
  $\tau$  denotes the mean
 time that it takes for a MT to rejoin the group of $n_+$ pushing MTs after 
 undergoing an individual catastrophe followed by rescue and force-free 
 growth at a velocity $v_+(0)$ that is larger  than the  velocity 
 $v_{+}({F}/{n_{+}})$ of the 
 leading MTs under force (see Fig.\ \ref{fig:tau}).
After a rescue time $1/\omega_r$ the trailing MT has to 
 ``catch-up'' a distance $[v_{+}({F}/n_{+})+v_-]/\omega_r$ to the 
leading MTs, which kept growing with velocity $v_{+}({F}/n_{+})$.
Given  a velocity difference $v_{+}(0)-v_{+}({F}/n_{+})$ to the 
leading MTs under force, this requires a time 
 \begin{equation}
  \tau \approx 
    \omega_r^{-1}\left[ 1+
      (v_{+}({F}/n_{+})+v_{-})/(v_{+}(0)-v_{+}({F}/n_{+}))   \right]
 \label{eq:tau}
 \end{equation}
 which is larger than the bare rescue time $1/\omega_r$.
This results in a  modified mean-field equation for $\langle n_{+}\rangle$,  
\begin{equation}
 d{\langle n_{+}\rangle}/dt =
  - \omega_c\left(\langle F\rangle/\langle n_{+}\rangle \right)
        \langle n_{+} \rangle
    + \langle \tau \rangle^{-1}\left(N-\langle n_{+}\rangle \right)
 \label{eq:avgn2}
\end{equation}
 where we have to apply the mean field averaging also 
 to $\langle \tau \rangle$ 
 in Eq.\  (\ref{eq:tau}):
 \begin{equation}
 \langle \tau \rangle \approx 
    \omega_r^{-1}\left[ 1+
      \frac{v_{+}(\langle F \rangle/\langle n_{+} \rangle)+v_{-}}
      {v_{+}(0)-v_{+}(\langle {F} \rangle/\langle n_{+} \rangle )}   \right].
 \label{eq:tau2}
 \end{equation}
Typically $\langle \tau \rangle$ is by a factor of 10 larger than
 the bare rescue time $1/\omega_r$,

\subsection{Limit cycle oscillations and absence of bifurcations}

For the further analysis of the mean-field dynamics
it is advantageous to introduce  new variables, 
the average force per MT  $f$ 
and the average fraction $\nu_+$ of pushing MTs,
\begin{equation}
f\equiv \langle F \rangle/N,~~~\nu_{+}\equiv \langle n_{+}\rangle/N
\end{equation}
 with 
$\langle F \rangle /\langle n_+ \rangle = f/\nu_+$.
Using these variables, 
the  mean-field equations  become 
\begin{eqnarray}
   d\nu_+ /dt &=&
  -  \nu_+\omega_c\left(f/\nu_+ \right) 
    + \left(1-\nu_+ \right)/\langle \tau \rangle
 \label{eq:avgnu}\\
  d{f}/dt &=& k 
     v_+\left(f/\nu_+ \right) /N
\label{eq:avgf}
\end{eqnarray}
We first discuss the nullclines of $f$ and $\nu_+$, i.e., 
the contours  in the $f$-$\nu_+$ plane along which
 $df/dt=0$ and $d\nu_+/dt=0$ is satisfied, respectively.

\begin{figure}
\begin{center}
   \epsfig{file=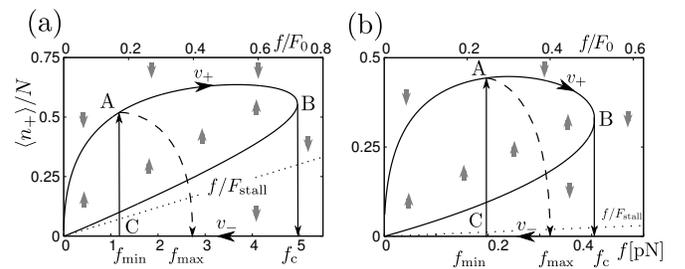,width=0.48\textwidth}
 \caption{\label{fig:meanfield}
Nullclines of the mean field equations for (a) the Flyvbjerg model 
 and (b) the Janson model for
     $\omega_{\text{on}}=70\;{\rm s}^{-1}$,  $\omega_r=0.05\;{\rm s}^{-1}$,
     and  $k=10^{-7}\;{\rm N/m}$ (soft barrier). 
 The  
 nullclines of  $\nu_+\equiv \langle n_+\rangle/N$ are solutions of 
 Eq.\ (\ref{eq:steady_nu}) and assume a loop shape as a 
 function of $f\equiv \langle F \rangle/N$.
 Gray arrows indicate the mean-field flow of $\langle n_+\rangle$.
The nullcline for $f$ is a 
  straight line  $\nu_+ = f/F_{\rm stall}$. 
 The critical force is 
  $f_c=F_c/N \simeq 3\;{\rm pN}$ for the Flyvbjerg model (a) and 
  $f_c \simeq 0.4\;{\rm pN}$ for the Janson model (b).
   Black arrows indicate the  stable mean-field limit cycle (see text);
  dashed line indicates   effect of 
   stochastic fluctuations.
}
\end{center}
\end{figure}

The nullclines of $f$ require 
 $v_+(f/\nu_+)=0$,  which leads to a straight line, 
 \begin{equation}
 f= \nu_+ F_{\rm stall}
\label{eq:steady_f}
\end{equation}
 in the $f$-$\nu_+$ plane, 
 where the slope is given 
 by  the single-MT stall force 
 $F_{\rm stall}= F_0\ln(\omega_{\text{on}}/\omega_{\text{off}})$
 [see Fig.\ \ref{fig:meanfield}(a)].

The nullclines  of $\nu_+$ are given by 
 \begin{equation}
  0 =  g(f,\nu_+) \equiv -\nu_+  \omega_c(f/\nu_+) + 
     \frac{1}{\langle \tau \rangle(f/\nu_+)} \left(1-\nu_+ \right)
 \label{eq:steady_nu}
\end{equation}
and are  {\em independent} of $N$. 
The shape of the  nullclines depends on the functional dependence 
of the catastrophe rate on the force and, thus,  on the catastrophe 
model. We will first focus on the  Flyvbjerg catastrophe model, for 
which the nullclines of $\nu_+$ have a 
characteristic loop shape, as shown in Fig.\  \ref{fig:meanfield}, 
which  exhibits  two solution branches:
a {\em stable} upper branch $\nu_{+,u}$
  corresponding to a collectively growing state with 
  $\langle n_+\rangle=N\nu_{+,u}$  pushing MTs and an 
 {\em unstable}  lower branch $\nu_{+,u}$.
 For a soft barrier, the  $\langle n_+ \rangle$ dynamics is fast, and the
 force increases slowly during collective growth, while 
 $\nu_+=\nu_{+,u}$ is tracing the stable upper branch 
  of the nullcline. 
The force per MT can increase up to a critical value $f_c$
 (with a corresponding value $\nu_c$ for $\nu_+$), where 
stable and unstable branch join and  where  the nullcline has vertical 
slope $df/d\nu_+=0$ in the $f$-$\nu_+$ plane.  
The critical  force $F_c = f_cN$ represents the maximal 
 load force for which  the MT ensemble can maintain a stable
 state of collective growth: for $\langle F \rangle>F_c$ the number
 $\langle n_+\rangle$ of pushing MTs has to flow spontaneously
  to a state  $\langle n_{+}\rangle=0$.

The critical force $f_c$ per MT can be obtained from 
two conditions: (i) the nullcline equation 
$g(f,\nu_+)=0$, i.e., Eq.\ (\ref{eq:steady_nu}), and
(ii) taking a total derivative 
with respect to $\nu_+$ and using the condition of a vertical slope 
$df/d\nu_+=0$ we arrive at the second 
condition $\frac{\partial}{\partial \nu_+}g(f,\nu_+)=0$.
Because $\omega_c(F)$ increases exponentially above $F_0$,
see Eq.\ (\ref{omegacv+}), 
$d\omega_c/dF \sim \omega_c/F_0$ is a good approximation.
The effective rescue time  $\langle \tau \rangle = 
\langle \tau \rangle(f/\nu_+)$ has a much weaker
force dependence, which we neglect. These approximations give 
\begin{eqnarray}
  0 
  &\approx&
   - \omega_c(f/\nu_+) \left(1-\frac{f}{\nu_+ F_0} \right)  -
     \frac{1}{\langle \tau \rangle(f/\nu_+)}
\nonumber\\
   \frac{f}{\nu_+ F_0} &\approx & 1+ \frac{1}{\omega_c(f/\nu_+)\langle \tau
     \rangle(f/\nu_+)} 
  \label{eq:fc2}
\end{eqnarray}
It turns out that (for both the Flyvbjerg and the Janson 
catastrophe models) 
$\omega_c \langle \tau \rangle  \ge 1$ holds over the entire range of forces.
In order to estimate $f_c$, we
 assume $\omega_c \langle \tau \rangle \gg 1$.  This leads to an  
estimate $\frac{f}{\nu_+} \approx F_0$ in Eq.\ (\ref{eq:fc2}),
which can be used in the arguments of $\omega_c$ and $\langle \tau \rangle$. 
Solving the Eqs.\ (\ref{eq:steady_nu}) and (\ref{eq:fc2}) 
for $f_c$ and $\nu_c$ we  find analytical estimates,
\begin{eqnarray}
   f_c &\approx & F_0 \frac{1}{\omega_c(F_0)\, \langle \tau\rangle(F_0)} 
   \label{eq:fc3}\\
  \nu_c &\approx& \frac{1}{1+\omega_c(F_0)\, \langle \tau\rangle(F_0)}.
  \label{eq:nuc}
\end{eqnarray}
According to Eqs.\ (\ref{eq:steady_f}) and (\ref{eq:steady_nu}),
the nullclines for $f$ and $\nu_+$ and, thus, 
the critical values $f_c$ and $\nu_c$ are strictly {\em independent} of 
$N$.  Therefore, the critical total force 
$F_c=Nf_c$ has to be  strictly {\em linear} in the number of MTs.
The critical force is the maximal polymerization force that 
can be generated during polymerization in the presence of 
rescue events. 
For a soft barrier ($N\gg \alpha$),
rescues thus lead to a significant increase in the maximal polymerization
force with a linear $N$ dependence 
compared to the  logarithmic dependence derived above in the absence 
of rescue.
Moreover, the estimate (\ref{eq:fc3}) for $f_c$  predicts an  increase 
of the generated force 
with the on-rate  $\omega_{\text{on}}$ because this increases $v_+$ and, 
thus,  reduces  $\omega_c$
 and an increase with the rescue rate $\omega_r$ because this decreases 
$\langle \tau \rangle$ [see Eq.\ (\ref{eq:tau})].

In order to  analyze the system for 
fixed points, we compare the lower branch $\nu_{+,l}$ of the 
nullcline  of $\nu_+$  with the  
nullcline (\ref{eq:steady_f}) of $f$.
The lower branch is governed by the exponential increase 
$\omega_c(F)\sim \omega_c(0)\exp(cF/F_0)$  with force
(with $c=2/3$ in the Flyvbjerg and $c=1$  in the Janson catastrophe model)
resulting in $\omega_c(0)\exp(cf/\nu_+ F_0) \sim 1/\langle \tau \rangle\nu_+$
 or 
$f/\nu_+ \approx F_0 \ln(1/\omega_c(0)\langle \tau \rangle\nu_+) /c$. 
This is always at {\em lower} forces than 
the nullcline [Eq.\ \ref{eq:steady_f})] of $f$
because $F_{\rm stall} = F_0 \ln(\omega_{\text{on}}/\omega_{\text{off}})\gg 
  F_0 \ln(1/\omega_c(0)\langle \tau \rangle \nu_+) /c$. 
This inequality can be violated only at very high rescue rates $\omega_r$
giving rise to a small $\langle \tau \rangle$.
 We obtained that  $\omega_r\gg 1/s$ is necessary to obtain a fixed point. 
Only if this fixed point exists {\em and} is stable, it can undergo 
a Hopf bifurcation on lowering the rescue rate.  
We conclude that, 
for realistic parameter values $\omega_r\sim  0.05\;\rm{s}^{-1}$, we are always
far from  a Hopf bifurcation.

The system rather {\em oscillates} in a {\em stable limit cycle}: 
After rescue 
[A in  Fig.\ \ref{fig:meanfield}(a)], the
 pushing force $f$ increases with  the MT growth velocity
because of $\dot{f}= k v_+/N$, while
$\nu_+=\nu_{+,u}$  is tracing the stable branch 
 of the nullcline. 
At the critical  force level $f_c$, 
a  collective catastrophe occurs [B in  Fig.\ \ref{fig:meanfield}(a)],
where the ensemble is quickly driven to  collective shrinking 
with $\langle n_+\rangle =0~ \mbox{or}~1$ and $\dot{f} = -kv_-/N$.

During shrinking  the force level is reduced until
an individual  rescue event can initiate  collective rescue at a  force
$F_{\rm min}$ [C in  Fig.\ \ref{fig:meanfield}(a)]. During rescue 
$\langle n_+\rangle$ increases  quickly back to its stable fixed point 
value [A in  Fig.\ \ref{fig:meanfield}(a)] closing the limit  cycle. 

The collective rescue force $F_{\rm min}$ can be calculated from the 
condition that  the lower unstable branch of fixed points 
given by  Eq.\ (\ref{eq:steady_nu}) intersects the line 
$\langle n_{+}\rangle=1$,
leading to the condition
\begin{equation}
 N=  \omega_c(F_{\rm min})
    \langle \tau \rangle(F_{\rm min}) + 1.
\label{Fmin}
\end{equation}
Collective rescue typically happens at rather small force 
$F_{\rm min}\ll F_0$ such that 
 we find  an essentially linear  $N$ dependence
 $F_{\rm min}\sim  N +\mathcal{O}(1)$.

The collective mean-field dynamics thus oscillates
  between forces $F_{\text{min}}$ and
$F_c$. The resulting time-averaged polymerization force 
\begin{equation}
F_{\text{s,N}} = \overline{\langle F \rangle} \approx (F_{\text{min}}+F_c)/2
\label{avF}
\end{equation}
  is also {\em  linear} in $N$.
This is  in agreement with the  simulation results 
[see Figs.\ \ref{fig:force}(a) and\ \ref{fig:force}(c)]. 
Because $F_c\gg F_{\text{min}}$ the result
$F_c \approx N F_0 /(\omega_c(F_0) \langle \tau\rangle(F_0))$
from Eq.\ (\ref{eq:fc3}) determines the dependence 
of the polymerization force $F_{\text{s,N}}$
on the on-rate  $\omega_{\text{on}}$ and   the rescue rate $\omega_r$.
 The estimate for $F_c$  predicts an  increase 
of the generated force 
with the on-rate  $\omega_{\text{on}}$ because this increases $v_+$ and, 
thus,  reduces  $\omega_c$.
 For the  velocity dependence (\ref{omegacv+}) and  assuming 
$v_+ \propto \omega_{\text{on}}$ 
(for $\omega_{\text{on}}\gg \omega_{\text{on}}$),  Eq.\ (\ref{eq:fc3})
gives 
$F_c \propto \omega_{\text{on}}^{2/3}$  for the Flyvbjerg catastrophe model 
(and $F_c \propto \omega_{\text{on}}$  for the Janson catastrophe model),
which is in qualitative 
 agreement with the simulation result of an approximately 
 linear  increase of the 
polymerization force with the on-rate $\omega_{\text{on}}$ in
 Figs.\ \ref{fig:force}(a) and\ \ref{fig:force}(d).
From  the result (\ref{eq:fc3}), we also predict 
 an increase of the  polymerization force
with the rescue rate $\omega_r$ because this decreases 
$\langle \tau \rangle$.
The   pronounced increase of  $F_{\text{s,N}}$
 with the on-rate  $\omega_{\text{on}}$ demonstrates 
 that for an MT ensemble, the  polymerization force can be 
sensitively regulated 
by changing the concentration of available monomers.
We also find 
the collective stall force $F_{\text{s,N}}$ always remains much smaller than 
the $N$-fold single-MT stall force, $F_{\text{s,N}}\ll N F_{\rm stall}$
[see Figs.\ \ref{fig:force}(a) and\ \ref{fig:force}(c)]
in contrast to force-sharing 
filaments  without dynamic instability, 
where  $F_{\text{s,N}}= N F_{\rm stall}$ holds exactly \cite{Krawczyk2011}.
A further confirmation  of the   mean-field theory is provided by 
simulation results for  the time-averaged 
 pushing fraction
   $\overline{\nu_{+}}= \overline{\langle n_+ \rangle}/N$
in Figs.\ \ref{fig:force}(c) and \ \ref{fig:force}(f). Mean field results for $\nu_+$ evaluated 
using the nullcline equation (\ref{eq:steady_nu}) for $f=F_{\text{s,N}}/N$
show good agreement with the simulations results.

The oscillatory limit cycle dynamics, which gives rise to 
collective catastrophe and rescue oscillations,  is 
robust against perturbations
because the system is far from a bifurcation for realistic 
rescue rates. Only for very high 
rescue rates $\omega_r\gg 1/s$, does  a stable fixed point 
exist, which becomes unstable in a Hopf bifurcation on 
lowering the rescue rate.

Similar collective catastrophes and rescues are also observed 
in {\em in vitro} bulk polymerization experiments
   \cite{Carlier1987,Marx1994}. In these experiments {\em many} MTs
   synchronously polymerize in a solution with  GTP-tubulin
   concentration $c_{\text{\text{GTP}}}$.  All MTs share the available
   concentration  $c_{\text{\text{GTP}}}$ and grow with  a velocity
   $v_{+}(c_{\text{GTP}})$, which decreases if GTP-tubulin is consumed. 
  Here, collective catastrophes and rescues are caused by 
   sharing the   concentration
  $c_{\text{\text{GTP}}}$ of available GTP-tubulin, resulting 
 in  similar collective oscillations   as force-sharing induces in 
the present system.

Finally, we want to note that the collective dynamics 
for $N\gg 1$ that we described here differ markedly 
from the dynamics of a {\it single} MT ($N=1$) \cite{Zelinski2012}. 
For a single MT rescue does not happen 
at a particular force level $F_{\text{min}}$ 
 but after an average time $1/\omega_r$ set by the individual rescue rate.
The resulting $N=1$ mean-field equation for the average force
$\langle F \rangle$ is 
 $v_{-}/\omega_r=v_{+}(\langle F \rangle)/\omega_c(\langle F \rangle)$
\cite{Zelinski2012} and 
equals shrinking and growing distance between individual rescue and
catastrophe events.

\subsection{Robustness with respect to catastrophe models}

An essential requirement for 
the existence of   an oscillatory limit cycle is 
the loop shape of the  nullclines  of  $\langle n_+ \rangle$
according to the stationary mean-field equation  (\ref{eq:steady_nu})
(see Fig.\ \ref{fig:meanfield}).
Results presented so far have been derived from the Flyvbjerg model. 
We obtain a very similar loop-shaped nullcline  also 
with the catastrophe model by Janson {\it et al.} [see Eq.\ (\ref{Janson}]. 
The condition for a loop-shape nullcline is the existence of a 
critical force  $F_c$, where the two solution branches 
of Eq.\ (\ref{eq:steady_nu}) merge in a point with a
vertical tangent. From Eq.\ (\ref{eq:steady_nu}), we can derive 
  the  necessary condition 
\begin{equation}
0<  \tau^{-1} <\omega_c(F) - F \frac{d}{dF} \omega_c(F)
\label{criticalFc}
\end{equation}
 for the existence of 
a critical force $F_c$. 
Therefore, we expect the same type of oscillatory limit cycle for 
 collective catastrophe and rescue oscillations for a large 
class of catastrophe models which meet
 the two conditions stated in Sec. \ref{sec_catmodels}:
(i) The catastrophe rate 
$\omega_c=\omega_c(v_+)$ is a 
function of the growth velocity only
and (ii) the resulting force dependence fulfills condition (\ref{criticalFc}),
which gives rise to a catastrophe rate increasing exponentially 
with force above the characteristic force $F_0$.
Whereas the Flyvbjerg and Janson catastrophe models and, more generally, 
all models with  $\omega_c \propto v_+^{-\varepsilon}$ 
($\varepsilon>0$) fulfill condition (\ref{criticalFc}),
it is violated for  the linear catastrophe model 
Eq.\ (\ref{linear}).

This explains  that the  mean-field result of an 
 oscillatory limit cycle is  robust with respect 
to variations of the catastrophe models: We expect
qualitatively similar 
behavior for all  catastrophe rates
$\omega_c(F)$, which are exponentially increasing with force above 
a characteristic force $F_0$, for example, in the standard 
catastrophe models
 by Flyvbjerg {\it et al.}  \cite{Flyv1994PRL,Flyv1996PRE}
or  by Janson {\it et al.}  \cite{Janson2003}.
This explains the robustness  observed in the simulation results 
 as shown in Figs.\  \ref{fig:trajectories} and \ref{fig:force}.

\begin{figure}
\begin{center}
   \epsfig{file=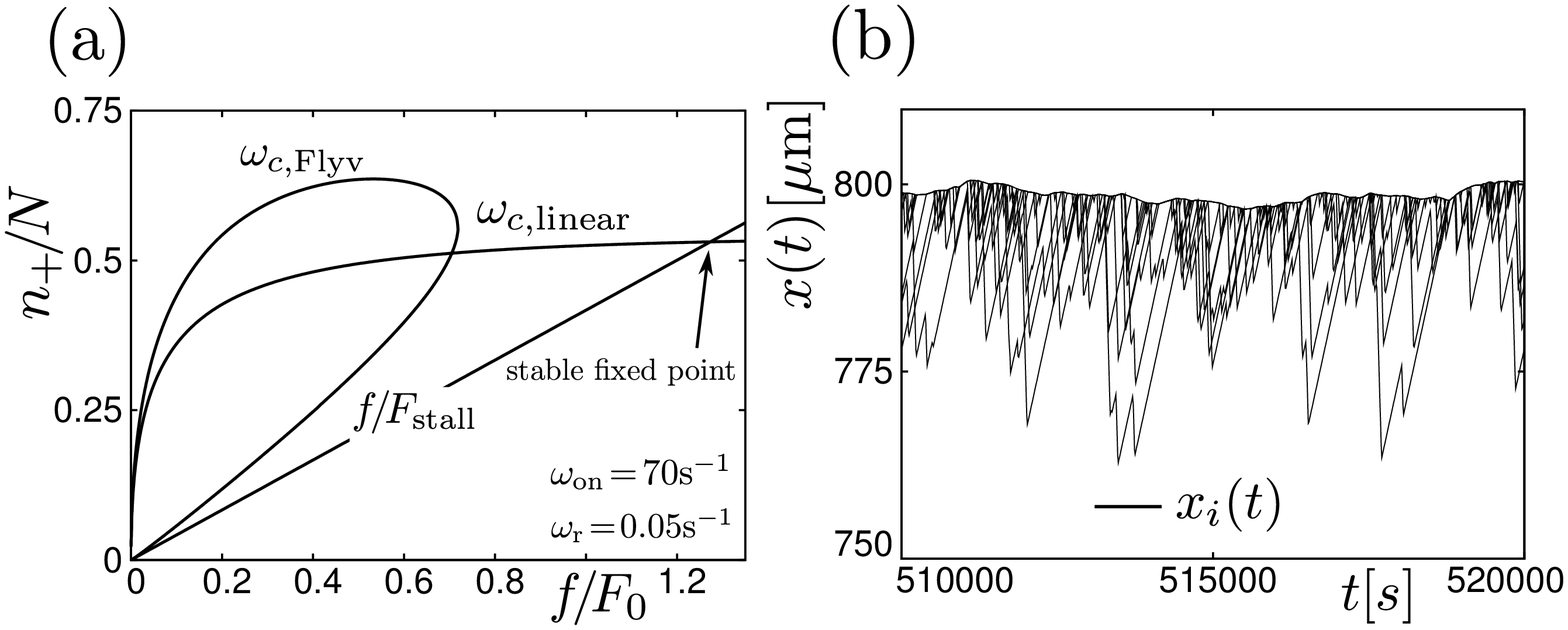,width=0.48\textwidth}
 \caption{\label{fig:linear}
(a) Nullclines of the mean field equations for the linear model
(\ref{linear}) with $\tilde{a} = 0.005\;{\rm s^{-1}}$ and 
   $\tilde{b} = 8 \times 10^5\;{\rm m^{-1}}$.
We use 
     $\omega_{\text{on}}=70\;{\rm s}^{-1}$,  $\omega_r=0.05\;{\rm s}^{-1}$,
     and  $k=10^{-7}\;{\rm N/m}$ (soft barrier). 
 The  nullcline for   $\nu_+\equiv \langle n_+\rangle/N$ 
is not loop-shaped. 
A stable fixed point exists  at the intersection with 
the nullcline for $f$, which  is the 
  straight line   $\nu_+ = f/F_{\rm stall}$. 
(b) Simulation trajectory of the 
  position $x(t)$ of all MTs as a function of time $t$ for $N=10$.
}
\end{center}
\end{figure}

The condition (\ref{criticalFc}) is violated for 
the linear catastrophe model 
Eq.\ (\ref{linear}). Accordingly, we do not expect to find 
an  oscillatory limit cycle  with collective catastrophe 
and rescue events. 
For this type of catastrophe model the nullclines are indeed no 
longer loop shaped, and the  mean-field theory rather predicts 
a {\em stable} fixed point [see Fig.\ \ref{fig:linear}(a)]. 
 Simulations confirm that collective catastrophe and rescue oscillations 
are absent for the linear catastrophe model, and we find a rather stationary 
 position $x$ of the obstacle and, thus, a stationary 
 polymerization force [see Fig.\ \ref{fig:linear}(b)].

\section{Stochastic fluctuations}

The dynamical mean-field theory explains all simulation results 
qualitatively. 
In order to obtain quantitative agreement with stochastic simulations,
we have to take into account that  the 
maximal  force $F_{\text{max}}$ for a collective catastrophe is
typically 
{\em smaller} than the critical-mean field force $F_c$ [see Figs.\
\ref{fig:force}(a) and\ 
\ref{fig:force} (c)]  because of 
additional stochastic fluctuations of $n_+$, which reduce 
the first passage time to a shrinking state $n_{+}=0$ [see Fig.\ \ref{fig:meanfield}].

Improved mean-field 
  results including this effect \cite{Zelinski}
agree quantitatively with full stochastic
simulations [see Fig.\ \ref{fig:force}(a) and\ \ref{fig:force}(c)].
We confirm the  {\em linear} $N$ dependence 
of the time-averaged force $F_{\text{s,N}}$ for small $N<10$, 
and  find a slightly 
stronger  than linear increase 
for larger $N$.  We also reproduce the 
approximately  increase of  $F_{\text{s,N}}$
 with the on-rate  $\omega_{\text{on}}$ 
[see Figs.\ \ref{fig:force}(b) and\ \ref{fig:force}(d)].

\section{Conclusion}

In cooperative force generation by an ensemble of $N$ MTs, 
the interplay between force-sharing and MT dynamic
 instability gives rise to a complex dynamics, which 
can be described in terms of collective 
catastrophe and rescue events.

We developed a dynamical mean-field theory [see Eqs.\ 
 (\ref{eq:avgnu}) and (\ref{eq:avgf})] which 
gives a quantitative description of 
 the  cooperative MT dynamics in terms of  two 
parameters, the mean force $\langle F \rangle$ and the 
mean number of pushing MTs $\langle n_+ \rangle$, in both the 
absence and presence of rescue events. 
Using this mean-field theory we identify 
 the relevant control parameters, such as 
tubulin on-rate, rescue rate, and MT number, and their influence on 
 force generation, and we  investigate 
 the  robustness against variations of the catastrophe model.
We validated the dynamical 
mean-field theory by  stochastic simulations of the  MT 
ensemble dynamics.

Our main findings are as follows.
In the absence of rescue events, the 
 maximal polymerization force before collective catastrophes 
 grows linearly with $N$  for 
small $N$ or  a stiff elastic barrier, in agreement with 
existing experimental  data \cite{Laan2008},
whereas it crosses over to a logarithmic dependence for larger $N$
or soft elastic barrier [see Eq.\ (\ref{Fmaxzero}) and Fig.\ 
\ref{fig:force_norescue}].
 This crossover
should be accessible in  experiments by 
varying the stiffness of optical traps.

In the presence of rescue events and 
for a {\it soft} elastic barrier, the dynamics 
becomes strongly modified:
Collective catastrophes and rescues lead to an oscillatory
stable limit cycle dynamics far from a Hopf bifurcation. 
These oscillations should be observable {\it in vitro}
in experiments such as in Ref.\ \cite{Laan2008} if the MT lengths
are sufficient to observe rescue events and if the stiffness of 
optical traps is reduced.
Moreover, {\it in vivo} the behavior of polarized MT ensembles 
can be explored, as has been shown in  Ref.\  \cite{Picone2010},
and our model predicts 
synchronized growth and shrinkage in oscillations 
if a polarized MT ensemble is growing against an elastic barrier
such as the cell cortex.

In the presence of oscillations, we have quantified 
the  maximal polymerization force $F_c=Nf_c$ in Eq.\ (\ref{eq:fc3})
and the time-averaged polymerization force $F_{\text{s,N}}$ in Eq.\
(\ref{avF}). Both forces  are  {\em linear} in $N$ [see  Figs.\ \ref{fig:force}(a) and\ \ref{fig:force}(d)], and the relevant force
scale is the force scale $F_0$, above which the MT growth velocity 
decreases exponentially and the MT catastrophe rate increases 
exponentially. 
The linear $N$ dependence of forces in the presence of rescue events is 
remarkable because we find an only logarithmic increase with $N$ 
in the absence of rescue events for soft barriers (see 
Fig.\ \ref{fig:force_norescue}).
Nevertheless, even the maximal polymerization force
 is  significantly smaller than the $N$-fold single-MT stall force.
This shows that MTs are 
not optimized with respect to force generation
because of their dynamic instability, even if they 
cooperate in an ensemble.
On the other hand, 
our analysis also shows that 
force generation in MT ensembles is  very sensitive to changes of 
system parameters related to the dynamic instability of MTs; 
in particular, it strongly increases 
with increasing  tubulin on-rate (and, thus, decreasing catastrophe rate)
 or increasing MT rescue rate
 (see Fig.\ \ref{fig:force}). 
The combination of both results suggests 
 that a MT ensemble is not efficient to generate high  forces
but that the dynamic instability in connection with the ensemble dynamics
allows us to efficiently regulate force generation through
several system parameters.
In the 
living cell, the on-rate can be changed by sequestering 
tubulin-dimers and catastrophe and  rescue rates are 
influenced, for example, 
by microtubule associated proteins 
\cite{Drechsel1992,Pryer1992,Dhamodharan1995}.

Our results also have 
implications for possible mechanisms which determine the mean length 
of the MT cytoskeleton, as it has been studied experimentally 
 in Ref.\  \cite{Picone2010}.
For a fixed stiffness $k$ of the opposing elastic force, 
the average force generated by the MT ensemble  corresponds 
to an average length of MTs.
The linear increase of the average force with the number $N$ of MTs
suggests that the MT length and, eventually, the cell size should 
exhibit a similar linear increase with the number of MTs in a
polarized MT cytoskeleton if the stiffness of the cell cortex 
remains unchanged.

{\it In vivo}, 
regulation mechanisms, which 
will involve the kinetics and spatial variation of 
concentrations of regulating proteins, 
will be relevant for cooperative 
MT dynamics and force generation.
The present work provides a theory to describe the cooperative 
dynamics arising from force sharing and its dependence
on various system parameters such as tubulin concentration
and rescue rates. This is a prerequisite  in order to 
explore spatial and temporal variations 
of these parameters in regulation mechanisms in future work.

\begin{acknowledgments}
We acknowledge support by  the Deutsche Forschungsgemeinschaft 
(KI 662/4-1).
\end{acknowledgments}


\end{document}